\documentclass[aps,showpacs,preprintnumbers,amsmath,amssymb]{revtex4}
\usepackage{dcolumn}
\usepackage[dvips]{epsfig}

\usepackage{float}
\usepackage{graphics,epsfig}
\usepackage{graphicx}
\usepackage{dcolumn}
\usepackage{bm}

\begin{document}
\baselineskip=0.8 cm
\title{\bf  Selected spherical photon orbits around a deformed Kerr black hole }

\author{
Changqing Liu$^{1}$\footnote{Electronic address:
changqingliu@ua.pt} Chikun Ding$^{1}$\footnote{Electronic
address:  Chikun Ding@huhst.edu.cn}, Jiliang
Jing$^{2}$\footnote{Electronic address: jljing@hunnu.edu.cn}}

\affiliation{1) Department of Physics, Hunan University of
Humanities Science and Technology, Loudi, Hunan 417000, P. R. China}

\affiliation{2) Department of Physics, and Key Laboratory of Low
Dimensional Quantum Structures and Quantum Control of Ministry of
Education, Hunan Normal University,  Changsha, Hunan 410081, P. R.
China}
\begin{abstract}
\baselineskip=0.6 cm
\begin{center}
{\bf Abstract}
\end{center}

In this paper, we investigate so-called spherical photon orbits around a deformed Kerr black hole with an extra deformation parameter.
The change in the azimuth $\Delta\varphi$  and the angle of dragging of nodes per revolution $\Delta\Omega $ of a complete latitudinal oscillated orbit is  calculated analytically. Finally, representative six examples orbits are plotted
out to illustrate how spherical photon orbits look like and exhibit some interesting behavior. Especially, this spherical photon orbits are absent in circular
orbits and different from the Kerr case.

\end{abstract}

\pacs{04.70.Bw,04.20.-q, 04.80.Cc} \maketitle
\newpage
\section{Introduction}
The investigation of null geodesic motion can
reveal significant features of a curved spacetime. Especially, there are unstable and stable photon
orbits around the compact objects. The unstable photon
orbits define the boundary between
capture and non-capture of across-section of light rays of black hole such as the shadow in lensing images\cite{sha2,sha3,sha4}, on the other
hand, the stable photon have directly link with the optical appearance\cite{sha5} of the thin accretion disk\cite{Luminet} and chaotic scattering in lensing around of hairy black holes and spacetime instabilities\cite{sw,fpos2,fpos3}. These fundamental photon orbits have an interested invariant structures around dynamical
systems and compact objects\cite{BI,mingzhi0}. In more specifically, the so-called spherical photon orbits\cite{Wilkins,teo}--orbits with constant coordinate radii that are not confined to the equatorial plane, have rich orbital structure, i.e. periodic orbit of the  longitudinal motion of particles. This orbits can further reveal the feature of black holes.

Spherical timelike orbits around the Kerr black hole was firstly proposed by Wilkins\cite{Wilkins}. An explicit example of spherical timelike orbits was plotted with numerical integration in paper\cite{Goldstein}. The extension to
the case of the charged Kerr Newman black hole was considered in Ref \cite{Johnston}. Further more,  an example of a
non-spherical timelike orbit around the Kerr black hole was obtained by numerical integration in Stoghianidis\cite{Stoghianidis} 's work. However, there has been less work done on spherical photon orbits. Early examples of spherical
photon orbits in the hyper-extreme Kerr space-time were illustrated in \cite{Schastok} and offered a tantalizing hint as to how
spherical photon orbits might look like. In Teo's paper\cite{teo}, several representative latitudinal oscillations photon orbits, including a zero-angular momentum photon orbit and one with non-fixed azimuthal
direction, was plotted to illustrate how the spherical photon orbits look like. These orbits exhibit a variety of interesting behavior that are absent in circular orbits.

Recently the LIGO \cite{gw1,gw2,gw3,gw4} and VIRGO collaborations reported the observation of gravitational-wave
signal corresponding to the inspiral and merger of two black holes. However, the current precision
of the experiment there remains some possibility for alternative
theories of gravity.  Konoplya and Zhidenko \cite{kz,RLs} have proposed a deformed Kerr black hole metric beyond general relativity through adding a static deformation,  which can be looked as an axisymmetric vacuum solution of a unknown alternative theory of gravity \cite{RLs}. This deformed Kerr black hole has three parameter, i.e., the  mass $M$, the rotation parameter $a$, and the deformation parameter $\eta$. $\eta$ describes the deviation
from the usual Kerr one and modifies sharply the structures of spacetime in the strong-field region. Moreover, research work about shadow of lensing\cite{mingzhi}, energy extraction\cite{fen}, strong gravitational lensing effect\cite{sy}, the iron line \cite{GKt02} and the quasi-periodic oscillations \cite{GKt01}  endorse that the geometry of
a real astrophysical black hole could be described by such a deformed Kerr metric.

In this paper, we shall focus on spherical photon orbits (with positive energy)
outside the event horizon of the deformed Kerr\cite{kz,RLs} black hole. We shall find it quite amazing that photons
can actually trace out such orbits around the deformed Kerr black hole, and also compared our result
with the Kerr case\cite{teo}. This may provide a possibility to test how astronomical black holes with
deformation parameter deviate from the Kerr black hole
 from the orbital motion aspect.

The paper is organized as follows: In Sec. II, we will derive the relevant geodesic equations of deformed Kerr black hole. In Sec. III, the conditions for the existence of
spherical photon orbits are considered. In Sec.IV,  the  expression for the change
in the orbit¡¯s azimuth for every oscillation in latitude. 3D orbit and the corresponding projective plane of selected
spherical photon orbits is plotted by numerical integration to illustrate how it differ from the Kerr case.
Finally, we end the paper
with a summary.

\section{The null geodesic equations in the deformed Kerr black hole }
The deformed Kerr metric obtained in Ref.\cite{kz,RLs} describes the geometry of a rotating black hole with the deviations from the Kerr one through adding an extra deformation. The deformed Kerr metric in the standard Boyer-Lindquist coordinates can be
expressed as
\begin{eqnarray}
\label{xy}
ds^{2} &=& -\bigg(1-\frac{2Mr^2+\eta}{r\rho^{2}}\bigg)dt^{2}
+\frac{\rho^{2}}{\Delta}dr^{2}+\rho^{2} d\theta^{2}+\sin^{2}\theta\bigg[r^{2}+a^{2}
+\frac{(2Mr^{2}+\eta)a^2\sin^{2}\theta}{r\rho^{2}}\bigg]d\varphi^{2}\\ \nonumber
&-&\frac{2(2Mr^2+\eta)a\sin^{2}\theta}{r\rho^{2}}dtd\varphi,
\end{eqnarray}
with
\begin{equation}
\Delta=a^{2}+r^{2}-2Mr-\frac{\eta}{r},\;\;\;\;\;\;\;\;\;\; \rho^{2}=r^{2}+a^{2}\cos^{2}\theta.
\end{equation}
where $M$, $a$ and $\eta$ denote the mass, the angular momentum and the deformation parameter of black hole, respectively. The deformation parameter $\eta$  describes the deviations from the Kerr metric. As the parameter $\eta$ vanishes, the metric reduces to the case of Kerr.In Ref \cite{kz,sy}, the condition for the existence of black hole horizon is analyzed in detail.
In the case $a<M$, the existence of black hole horizon require \cite{sy}
\begin{equation}
\label{hcz}
\eta\geq \eta_{c1}\equiv-\frac{2}{27}(\sqrt{4M^{2}-3a^{2}}+2M)^{2}(\sqrt{4M^{2}-3a^{2}}-M),
\end{equation}
while in the case $a>M$, it becomes $\eta>0$. When $\eta$ and $a$ lie in other regions, there is no
horizon and then the spacetime (\ref{xy}) becomes a naked singularity.
Thus, the value of $\eta$ determines the number and positions of black hole horizons.  These spacetime properties affect the propagation of photon and further changes  shadow of  deformed Kerr black hole.

The Hamiltonian of a photon propagation along null geodesics in a deformed Kerr black hole  can be expressed as
\begin{equation}
H(x_i, p_i) =\frac{1}{2}g^{\mu\nu}(x)p_{\mu}p_{\nu}=
\frac{\Delta}{2\rho^2}p_r^2+\frac{1}{2\rho^2}p_\theta^2+f(r,\theta,p_t,p_\varphi)=0 ,
\end{equation}
Since there exist two ignorable coordinates $t$ and $\phi$ in the above Hamiltonian, it is easy to obtain two conserved quantities $E$ and $L_{z}$ with the following forms
\begin{eqnarray}
\label{EL}
E=-p_{t}=-g_{tt}\dot{t}-g_{t\varphi}\dot{\varphi},\;\;\;\;\;\;\;\;\;\;\;\;\;\;
L_{z}=p_{\varphi}=g_{\varphi\varphi}\dot{\varphi}+g_{\varphi t}\dot{t},
\end{eqnarray}
which correspond to the energy and angular momentum
of photon moving in the deformed Kerr black hole. With these two conserved quantities, after tedious calculation, we obtain the null geodesic equation\cite{mingzhi0}.
\begin{eqnarray}
\label{tfc}
\dot{t}&=&E+\frac{(a^{2}E-aL_{z}+Er^{2})(2Mr^{2}+\eta)}{\Delta\rho^{2}r},
\\
\label{jfc}
\dot{\varphi}&=&\frac{aE\sin^{2}\theta(2Mr^{2}+\eta)
+a^{2}L_{z}r\cos^{2}\theta-L_{z}(2Mr^{2}
-r^{3}+\eta)}{\Delta\rho^{2}r\sin^{2}\theta},\\
\label{rfc}
\rho^{4}\dot{r}^{2}&=&R(r)=-\Delta[Q+(aE-L_{z})^{2}]+[aL_{z}-(r^{2}+a^{2})E]^{2},
\\
\label{thfc}
\rho^{4}\dot{\theta}^{2}&=&p_{\theta}^{2}=\Theta(\theta)=Q-\cos^{2}\theta
\bigg(\frac{L_{z}^{2}}{\sin^{2}\theta}-a^{2}E^{2}\bigg),
\end{eqnarray}
where the quantity $Q$ is the generalized Carter constant related to the constant of separation $K$ by $Q=K-(aE-L_{z})^2$.

We note that while these equations are concise and appealing in some
ways, during numerical integration they tend to accumulate error at
the turning points due to the explicit square roots in the $r$ and
$\theta$ equations, not to mention the nuisance of having to change
the signs of the $r$ and $\theta$ velocities by hand at every turning
point. Following the procedure in Ref \cite{Levin} , We will convert this equation into
 Hamiltonian formulation to avoid the numerical difficulties and smoothly plot the selected
spherical photon orbits.

 So we rewrite the Hamiltonian as following
 \begin{equation}
H(x_, p_i) =
\frac{\Delta}{2\rho^2}p_r^2+\frac{1}{2\rho^2}p_\theta^2
-\frac{R + \Delta\Theta}{2\Delta \rho^2}
\quad\quad ,
\label{niceham}
\end{equation}
with the help of Hamilton's
equations
\begin{equation}
\label{D-Eq} \frac{dx^i}{d \lambda} = \frac{\partial H}{\partial
p_i}  \, , \; \; \frac{dp_i}{d \lambda} = - \frac{\partial
H}{\partial x^i} \, ,
\end{equation}
for the non-zero  Hamiltonian formulation of the photon's motion become as
\begin{eqnarray}
\label{eom}
\dot{r} & =& \frac{\Delta}{\rho^2}p_{r}\label{geeoss1} , \\
 \dot{p}_{r} & = &
-\left (\frac{\Delta}{2\rho^2}\right )'p_{r}^{2} -
\left (\frac{1}{2\rho^2}\right )'p_{\theta}^{2} + \left (\frac{R +
 \Delta\Theta}{2\Delta\rho^2}\right )' \label{geeoss2}, \\
\dot{\theta}& = & \frac{1}{\rho^2}p_{\theta}
 ,\\
 \dot{p}_{\theta} & = &
-\left (\frac{\Delta}{2\rho^2}\right )^{\theta}p_{r}^{2} -
 \left (\frac{1}{2\rho^2}\right )^{\theta}p_{\theta}^{2} + \left (\frac{R +
 \Delta\Theta}{2\Delta\rho^2}\right )^{\theta}  \label{geeoss3},\\
  \dot{t} & = & \frac{1}{2\Delta\rho^2} \frac{\partial (R+\Delta\Theta)}{\partial
E}\label{geeoss4}, \\
\dot{\varphi} & = & -\frac{1}{2\Delta\rho^2}
  \frac{\partial(R +\Delta\Theta)}{\partial
L}, \label{geeoss5}
\end{eqnarray}
where the superscripts $'$ and $\theta$ denote differentiation with
respect to $r$ and $\theta$, respectively.
\section{Spherical Photon Orbits around a deformed Kerr Black Hole }
\begin{figure}[ht]
\begin{center}
\includegraphics[scale=1]{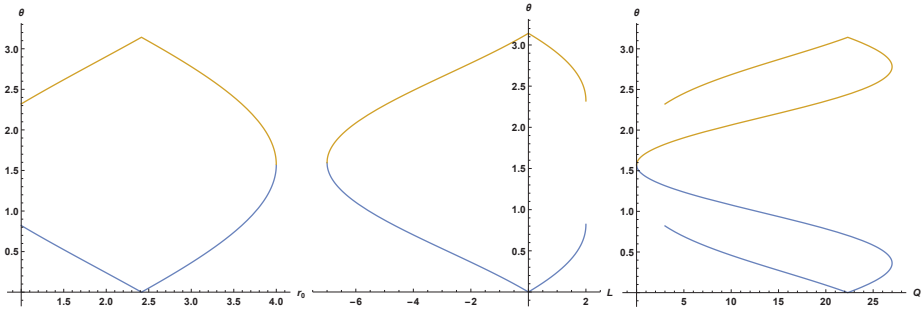}
\caption{Variation of $\theta$ range of latitudinal oscillations  with
the initial radius of $r_0$,  $L$, $Q$ in extreme Kerr black hole, here we set M=1 and E=1. \label{figure1}}
\end{center}
\end{figure}

In this section, we shall study in detail the properties of the spherical  photon orbit. The so-called spherical photon orbits--orbits with constant coordinate radii that are not confined to the equatorial plane and photon move as latitudinal oscillations in the permitted maximum $\theta$ angle\cite{Wilkins,teo}.  The radial equation of motion in the spherical photon orbits is unstable and satisfy
\begin{eqnarray}
\dot{r}=0,\;\;\;\;\; \text{and} \;\;\;\; \ddot{r}=0,
\end{eqnarray}
which yield
\begin{eqnarray}
\label{r}
R(r)&=&-\Delta[Q+(aE-L_{z})^{2}]+[aL_{z}-(r^{2}+a^{2})E]^{2}=0,
\\
\label{r1}
R'(r)&=&-4Er[aL_{z}-(r^{2}+a^{2})E]-[Q+(aE-L_{z})^{2}]
(-2M+2r+\frac{\eta}{r^{2}})=0.
\end{eqnarray}
For the unstable spherical orbits, we have
\begin{eqnarray}
\label{r2}
R''(r)=8E^{2}r^{2}-4E[aL_{z}-(r^{2}+a^{2})E]-2[Q+(aE-L_{z})^{2}]
(1-\frac{\eta}{r^{3}})>0.
\end{eqnarray}
Solving the two equation (\ref{r}) and (\ref{r1}), we find that for the spherical orbits motion of photon the reduced constants $\xi$ and $\sigma$
have the form
\begin{eqnarray}
\label{pj}
\xi&\equiv&\frac{L_{z}}{E}=\frac{2a^{2}Mr^{2}-a^{2}\eta+2\Delta r^{3}-2Mr^{4}-3\eta r^{2}}{a(2Mr^{2}-2r^{3}-\eta)},\\
\label{q}
\sigma&\equiv&\frac{Q}{E^{2}}=\frac{-r^{4}[(6Mr^{2}-2r^{3}+5\eta)^{2}-8a^{2}(2Mr^{3}+3\eta r)]}{a^{2}(2r^{3}-2Mr^{2}+\eta)^{2}}.
\end{eqnarray}
From Eq. (\ref{thfc}), we find that $\xi$ and $\sigma$ obey
\begin{eqnarray}
\label{fw}
\sigma-\xi^{2}\cot^2\theta+a^2\cos^{2}\theta\geq0.
\end{eqnarray}
If we set $u = cos\theta$,
when $\sigma$ is non-negative,  The physically allowed ranges for $u_0$ is given as
 \begin{eqnarray}
\label{fw11}
u_0^2=\frac{(a^2-\xi-\sigma^2)+\sqrt{(a^2-\xi-\sigma^2)^2+4a^2\sigma}}{2a^2}
\end{eqnarray}
As the initial radius of $r_0$ the spherical  photon orbit is given, The physically allowed angular momentum $L$, Carter constant $Q$,
 $\theta$ range is also determined. we take the extreme Kerr black hole as an example to illustrate the relationship between  $\theta$ and $r_0$, $L$, $Q$. in Fig. \ref{figure1}. photon can oscillate between $arccos(u_0)$ and  $arccos(-u_0)$, such orbits cross the equatorial plane repeatedly. all orbits either remain in the equatorial plane or cross it repeatedly. i.e. as the
 Carter constant $Q$ =0, photon orbit is entirely in the equatorial plane, while the
 Carter constant $Q$ =27, photon orbit is entirely in the $\theta$ direction. Especially, the zero angular momentum  photon orbit
 has reached the maximum $\theta$ value.

 Since the spherical photon orbit is latitudinal oscillations and periodicity. it is useful to have a measure of this periodicity. One possibility is to consider the change in azimuth $\Delta\varphi$  for a complete latitudinal oscillation of the orbit. It turns out to be possible to obtain an exact expression for the azimuth $\Delta\varphi$  for the photon orbit\cite{Wilkins,teo,Goldstein,Johnston},  as we now briefly describe the azimuth $\Delta\varphi$ by finding the connection between $\theta$ and $\varphi$ motions.

 If we set $w=u^2$, using Eqs. (\ref{jfc}) and (\ref{thfc}), we have
 \begin{eqnarray}
\label{thphi}
\frac{d\varphi}{dw}=\left(\frac{a(2Mr^2+\eta-a\xi~r)}{2\Delta r}+\frac{\xi}{2(1-w)}\right)\frac{1}{Y(w)},
\end{eqnarray}
where
 \begin{eqnarray}
\label{thphis}
Y(w)^2=\sigma~w-(\sigma+\xi^2-a^2)w^2-a^2w^3.
\end{eqnarray}
It would be useful to write the latter in the form
$-a^2w(w-w_+)(w-w_-)$, where $w_+$ are the positive and negative roots of $Y(w)^2$,
respectively. Then the change in
azimuth for one complete oscillation in latitude is
\begin{eqnarray}
\label{latitude}
\Delta\varphi=\frac{2a(2Mr^2+\eta-a\xi~r)}{\Delta r}\int_0^{w_+}\frac{dw}{Y(w)}+2\xi\int_0^{w_+}\frac{dw}{(1-w)Y(w)}.
\end{eqnarray}
These integrals can be evaluated using standard techniques to give
\begin{eqnarray}
\label{latitudeaas}
\Delta\varphi=\frac{4}{\sqrt{w_+-w_-}}\left(\frac{2Mr^2+\eta-a\xi~r}{\Delta r}K(\sqrt{\frac{w_+}{w_+-w_-}})+\frac{\xi}{a}\frac{1}{1-w_+}\Pi(\frac{-w_+}{1-w_+},
\sqrt{\frac{w_+}{w_+-w_-}}~)\right),
\end{eqnarray}
where $K(x)$ and $\Pi(\upsilon,x)$ are the complete elliptic integrals of the first and third kind, respectively.

\begin{figure}[ht]
\begin{center}
\includegraphics[scale=1]{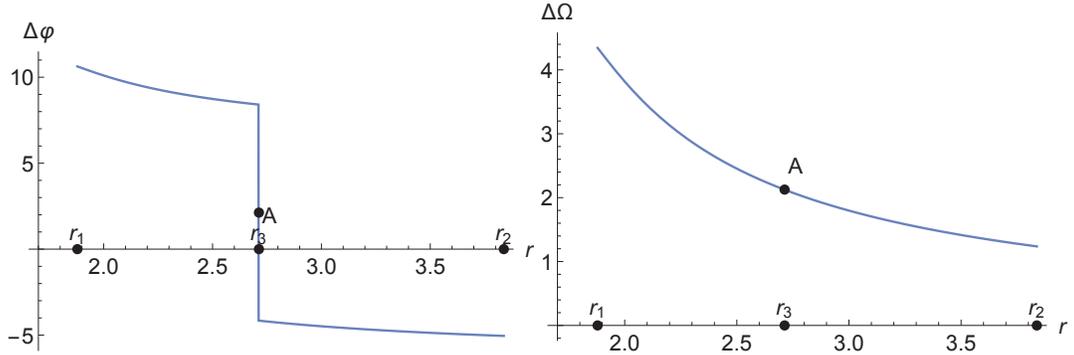}
\caption{Variation of the azimuth $\Delta\varphi$ and the angle of advance  $\Delta\Omega$ of the nodes  with
the value $r$ in  deformed Kerr black hole, $r_1$ and $r_2$ are the solution of
 the carter constant $\sigma=0$, $r_3$ is the solution of
 angular momentum $\xi=0$, the point $A$ is the exactly value $\Delta\varphi~(r_3)$ and  $\Delta\Omega$. Here we set $M=1$, $\eta=0.1$ and E=1. \label{figure2}}
\end{center}
\end{figure}

The specific dependence of the azimuth $\Delta\varphi$ on $r$ is shown in the left Fig.~\ref{figure2}.
When $r_1<r<r_3$, the orbits are prograde whenever $\Delta\varphi$ is positive, on the other hand, When $r_3<r<r_2$, the orbits are reprograde whenever $\Delta\varphi$ is negative ( $r_1$ and $r_2$ are the solution of
 the carter constant $\sigma=0$, $r_3$ is the solution of
  zero angular momentum $\xi=0$).  Notice that there is a  discontinuity at $r=r_3$( zero angular momentum $\xi=0$), the value $\Delta\varphi$ at exactly $r=r_3$  , given by the Point $A$, is exactly half-way between the upper limit $\lim_{r\rightarrow~r^+_3}\Delta\varphi$ and the lower limits $\lim_{r\rightarrow~r^-_3}\Delta\varphi$. It turns
out that there is a satisfying explanation for this behavior of 3D orbit of zero angular momentum photon in Fig. \ref{figure2}.

Following Winkins\cite{Wilkins}, we can define the ratio between the
frequencies in the $\varphi$ and $\theta$ direction by
\begin{eqnarray}
\label{feqs}
f=\frac{\upsilon_\varphi}{\upsilon_\theta}=\frac{|\Delta\varphi|}{2\pi}.
\end{eqnarray}
The angle of advance of the nodes(a point where a non equatorial orbit intersections the equatorial plane ) per nodal period  $\Delta\Omega$ \cite{Wilkins} is
\begin{eqnarray}
\label{feqssa}
\Delta\Omega=2\pi|f-1|.
\end{eqnarray}
We can find in the right of Fig.~\ref{figure2}, compared with the case of the azimuth $\Delta\varphi$,  the
change of the angle of advance of the nodes $\Delta\Omega$ is continuously.

Although each orbit that we are considering has a definite non-zero value for
$\Delta\varphi$, it is not guaranteed that the photon is moving in a fixed azimuthal direction at
every point of its orbit,  In fact, it follows from  Eq.(\ref{jfc}) that $\dot{\varphi}$ changes sign whenever
$u^2$ reaches the value
\begin{eqnarray}
\label{usb}
u_1^2&=&\frac{-2aMr^2+2\xi~Mr^2-Lr^3-a\eta+\xi\eta}{a(a\xi~r-2Mr^2-\eta)}\\\nonumber
&=&\frac{(6Mr^2-2r^3+5\eta)r^2}{a^2(2r^2(M+r)-\eta)}.
\end{eqnarray}
We define the value $r$ of satisfying  $\dot{\varphi}=0$ as $r_{\dot{\varphi}}$ and the corresponding angular momentum
 as $\xi_{\dot{\varphi}}$\cite{teo}. Note that when $r_3<r<r_{\dot{\varphi}}$  (corresponding to $\xi_{{\dot{\varphi}}}<\xi<0$), orbits with these parameters would therefore not be moving in a fixed azimuthal direction. An example of such an orbit will
also be given in the following section.
To compared with the result about the azimuth $\Delta\varphi$ and the angle of advance of the nodes $\Delta\Omega$ in the extreme Kerr black hole\cite{teo}, we
list the value of the azimuth $\Delta\varphi$ and the angle of advance of the nodes  $\Delta\Omega$ with the
given initial angular momentum $\xi$ in  deformed Kerr black hole in Table.~\ref{table1}. We find that for the given
 initial angular momentum $\xi$,  the presence of the deformation parameter $\eta$ decrease the value of azimuth $\Delta\varphi$ and the angle of advance of
the nodes  $\Delta\Omega$.
\begin{table}
\begin{center}
\caption{The value of the azimuth $\Delta\varphi$ and the angle of advance of the nodes  $\Delta\Omega$  with the
given initial  angular momentum $\xi$ in the deformed Kerr black hole. Here we set M=1.}
\begin{tabular}{c c c c c c c c}
  \hline
  \hline
  & $\eta$=0 & $\eta$=0.5 & $\eta$=1 & $\eta$=2  & $\eta$=3 & $\eta$=4\\
  \hline $\Delta\varphi_{\xi=0}$\;\; &  3.1761  & 2.4213 & 2.0420 &1.6253 & 1.3869 &1.2267 \\
 $\Delta\Omega_{\xi=0}$  & 3.1071 & 3.8619 & 1.9160 & 4.6579 &4.8963 & 5.0565  \\
  \hline $\Delta\varphi_{\xi=-1}$\;\; &  -3.7128  & -4.1351 & -4.3962 & -4.7203 &-4.9924 &-5.0647 \\
 $\Delta\Omega_{\xi=-1}$  & 2.5694 & 2.1481 & 1.8870 & 1.5629 &1.3607 & 1.2185  \\
   \hline $\Delta\varphi_{\xi=\xi_{\dot{\varphi}}}$\;\; &  -4.0728  &-4.3250 &-4.5101& -4.7688 & -4.9449 &-5.0747 \\
 $\Delta\Omega_{\xi=\xi_{\dot{\varphi}}}$  & 2.2104 & 1.9582 & 1.7731& 1.5144 &1.3382 & 1.2084  \\
   \hline $\Delta\varphi_{\xi=-6}$\;\; &  -4.7450  &-4.8296 &-4.9017 &-5.0191 & -5.1117 &-5.1874 \\
 $\Delta\Omega_{\xi=-6}$  & 1.5382 & 1.4535 & 1.3815 & 1.2641 &1.1714 & 1.0957\\
  \hline $\Delta\varphi_{\xi=1}$\;\; &  10.8428  & 9.0649 & 8.4874 &7.9519 & 7.6760 &7.4999 \\
 $\Delta\Omega_{\xi=1}$  & 4.5596 & 2.7817 & 2.20426 &1.6688 &1.3928& 1.2167  \\
   \hline $\Delta\varphi_{\xi=1.999}$\;\; &  159.418  & 9.4422 & 8.5860 & 7.9392 & 7.6396 &7.4570 \\
 $\Delta\Omega_{\xi=1.999}$  & 153.135 & 3.1591 &2.3028 & 1.6560 &1.3564 & 1.1738  \\
 \hline
  \hline\label{table1}
\end{tabular}
\end{center}
\end{table}

\begin{figure}[ht]
\begin{center}
\includegraphics[scale=1.1]{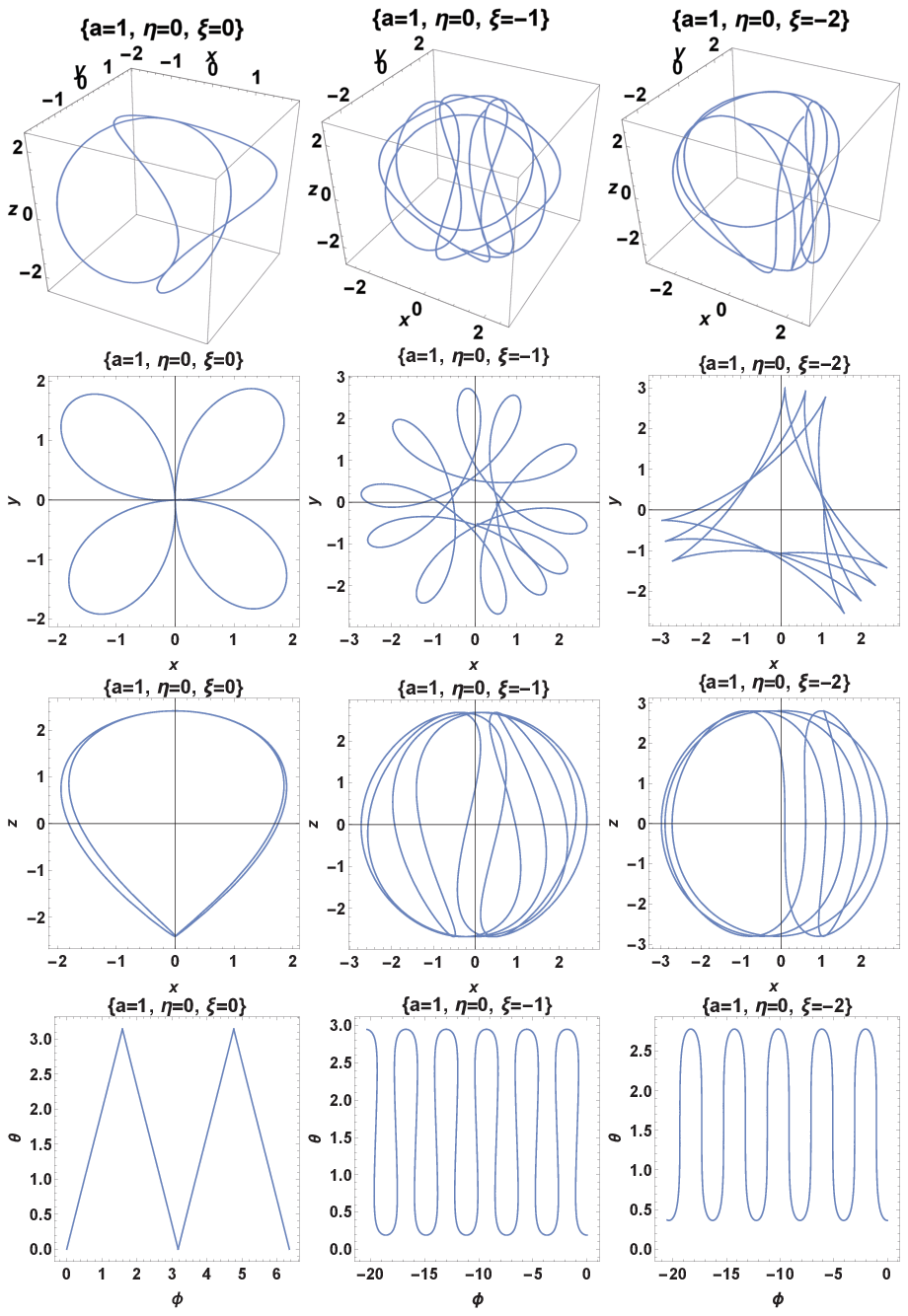}
\caption{The three-dimensional($x-y-z$) plane, the projective $x-y$ and $x-z$ plane, the $\theta-\varphi$ plane of the spherical photon orbits around the extreme Kerr black hole with the given initial angular momentum $\xi=0, -2, -1$ ,respectively.  Here we set $M=1$. \label{figure3}}
\end{center}
\end{figure}
\begin{figure}[ht]
\begin{center}
\includegraphics[scale=1]{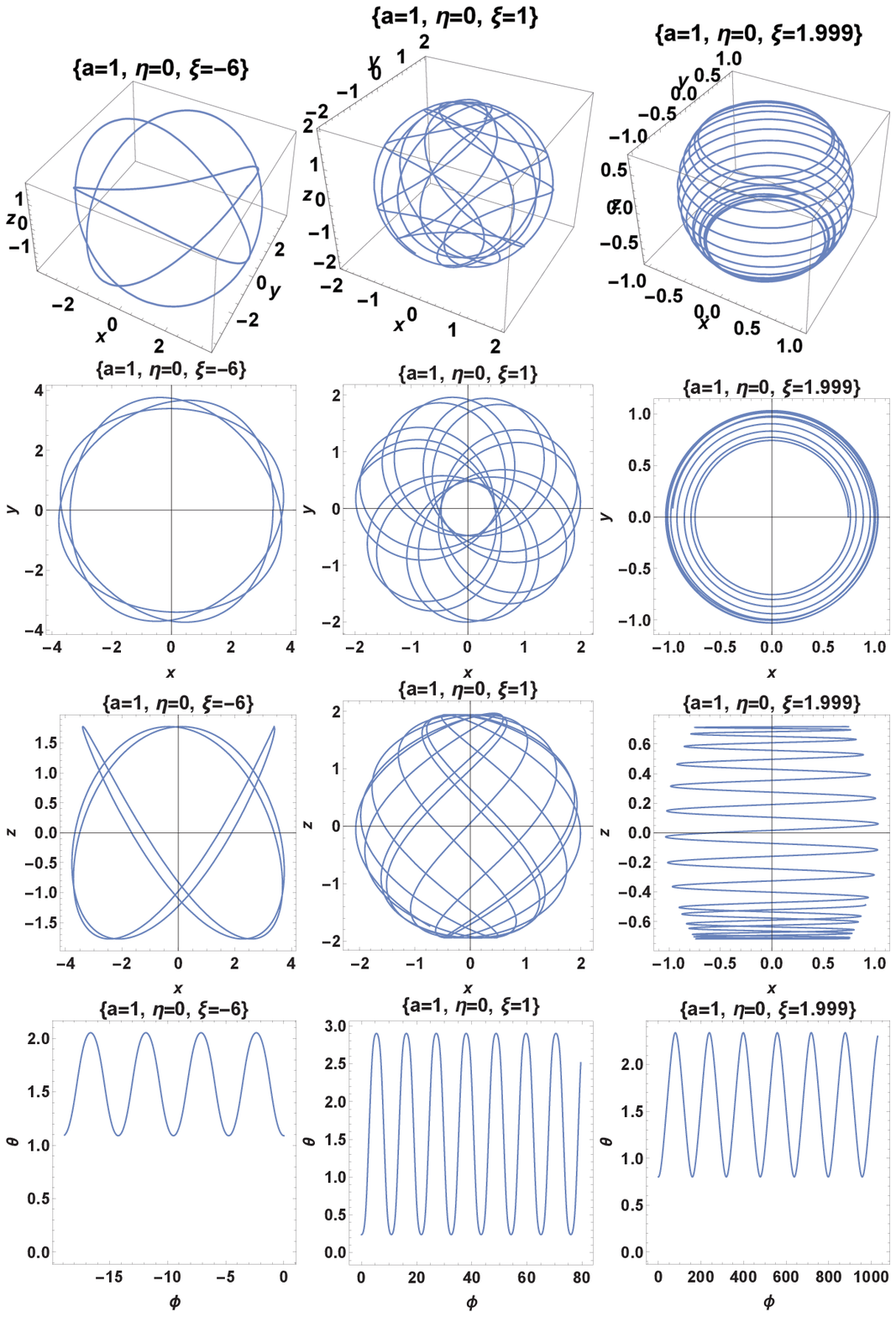}
\caption{The three-dimensional($x-y-z$) plane, the projective $x-y$ and $x-z$ plane, the $\theta-\varphi$ plane of the spherical photon orbits around the extreme Kerr black hole with the given initial angular momentum $\xi=-6, 1, 1.999$ ,respectively.  Here we set $M=1$. \label{figure4}}
\end{center}
\end{figure}
\begin{figure}[ht]
\begin{center}
\includegraphics[scale=0.8]{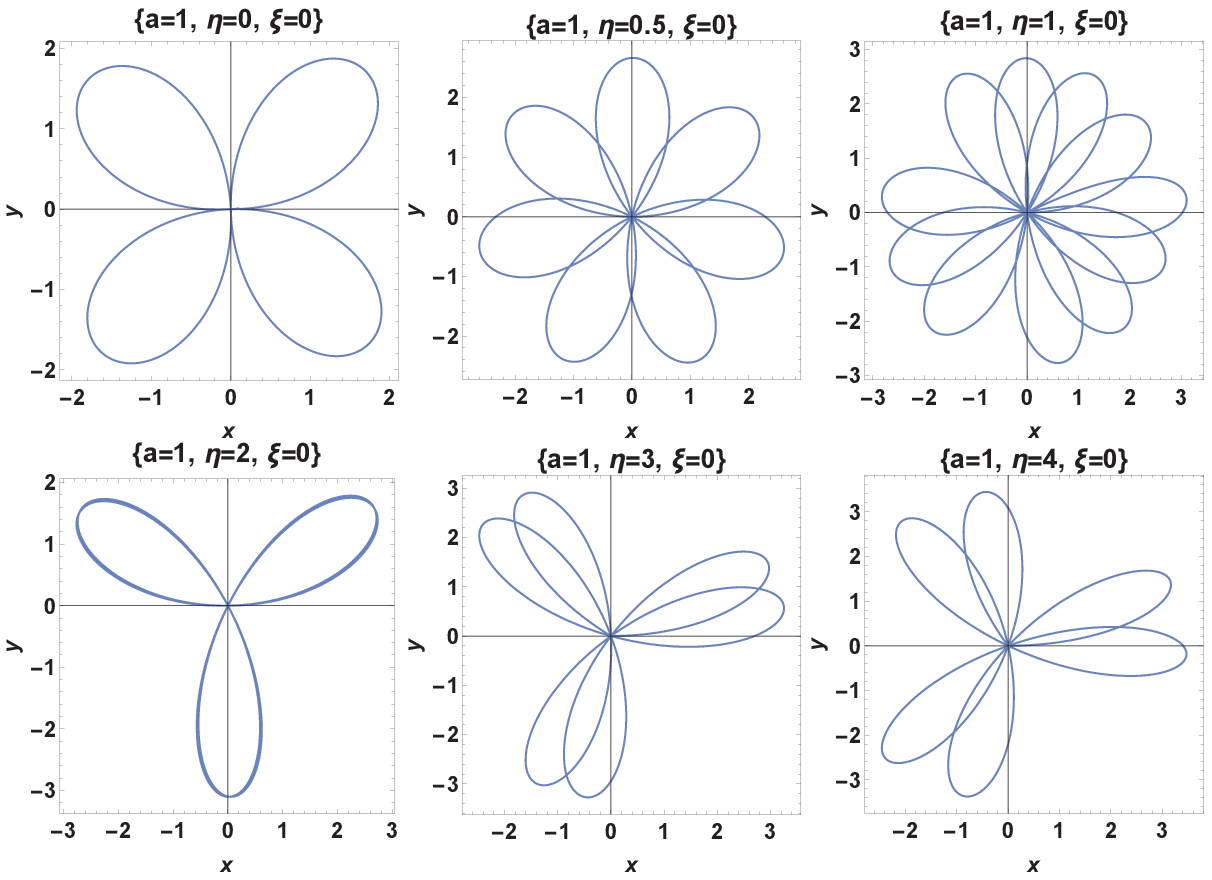}
\caption{ the projective  $x-y$ plane of the spherical photon orbits around the deformed Kerr black hole with the zero angular momentum $\xi=0$ for a different  deformation parameter $\eta$.  Here we set $M=1$. \label{fig1}}
\end{center}
\end{figure}
\begin{figure}[ht]
\begin{center}
\includegraphics[scale=0.8]{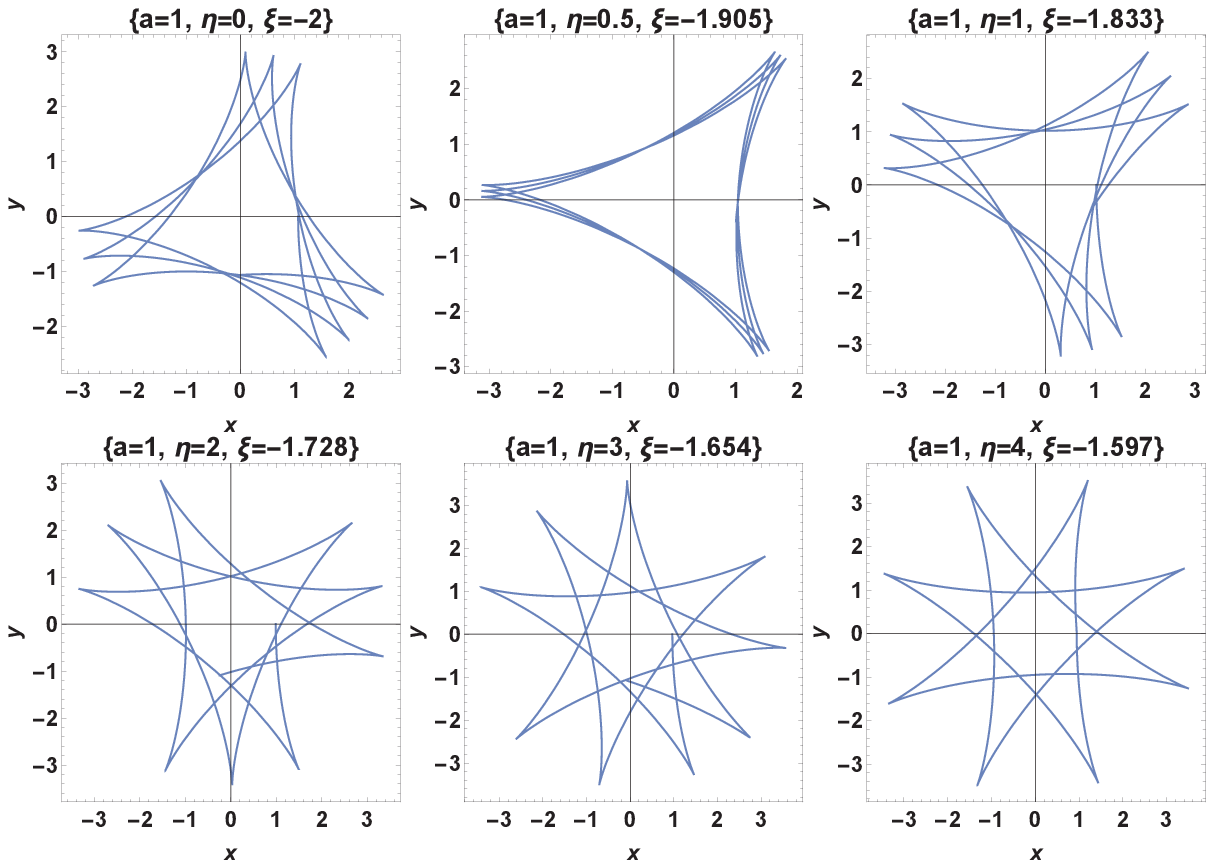}
\caption{The projective $x-y$ plane of the spherical photon orbits around the deformed Kerr black hole with the given initial angular momentum $\xi=\xi_{\dot{\varphi}}$ where satisfying $\dot{\varphi}=0$ for a different deformation parameter $\eta$. Here we set $M=1$. \label{fig2}}
\end{center}
\end{figure}\begin{figure}[ht]
\begin{center}
\includegraphics[scale=0.8]{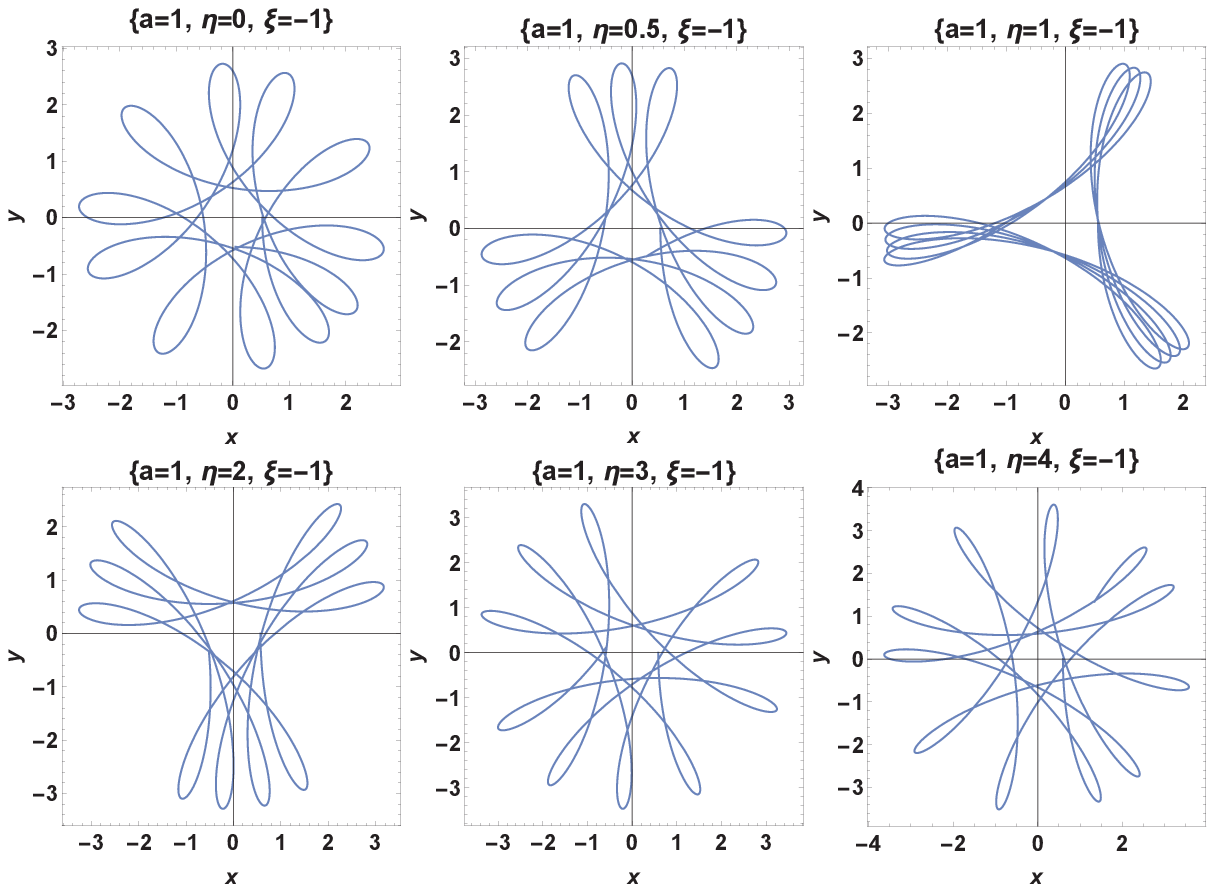}
\caption{The projective $x-y$ plane of the spherical photon orbits around the deformed Kerr black hole with the given initial angular momentum $\xi=1$ for a different deformation parameter $\eta$.  Here we set $M=1$. \label{fig3}}
\end{center}
\end{figure}
\begin{figure}[ht]
\begin{center}
\includegraphics[scale=0.8]{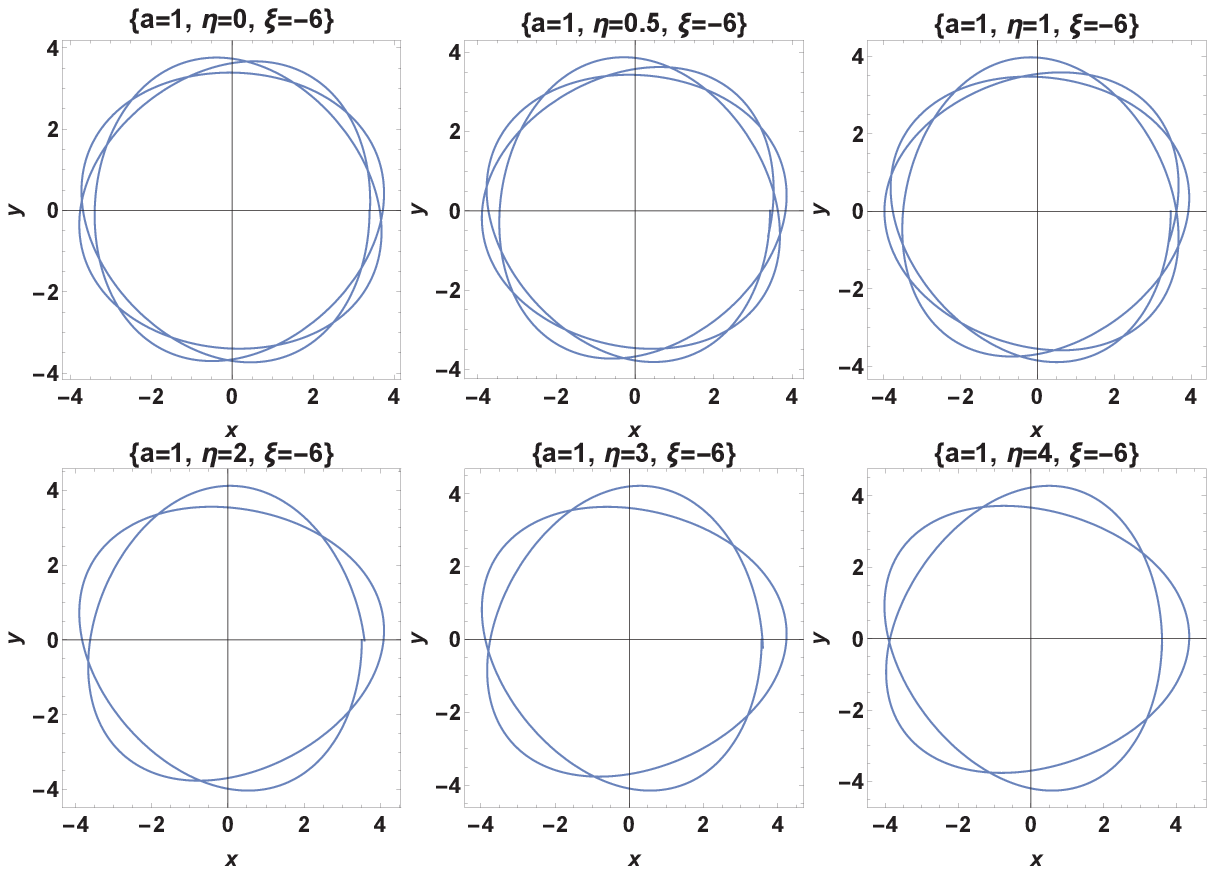}
\caption{The projective $x-y$ plane of the spherical photon orbits around the deformed Kerr black hole with the given initial angular momentum $\xi=-6$ for a different deformation parameter $\eta$..  Here we set $M=1$ \label{fig4}}
\end{center}
\end{figure}
\begin{figure}[ht]
\begin{center}
\includegraphics[scale=0.8]{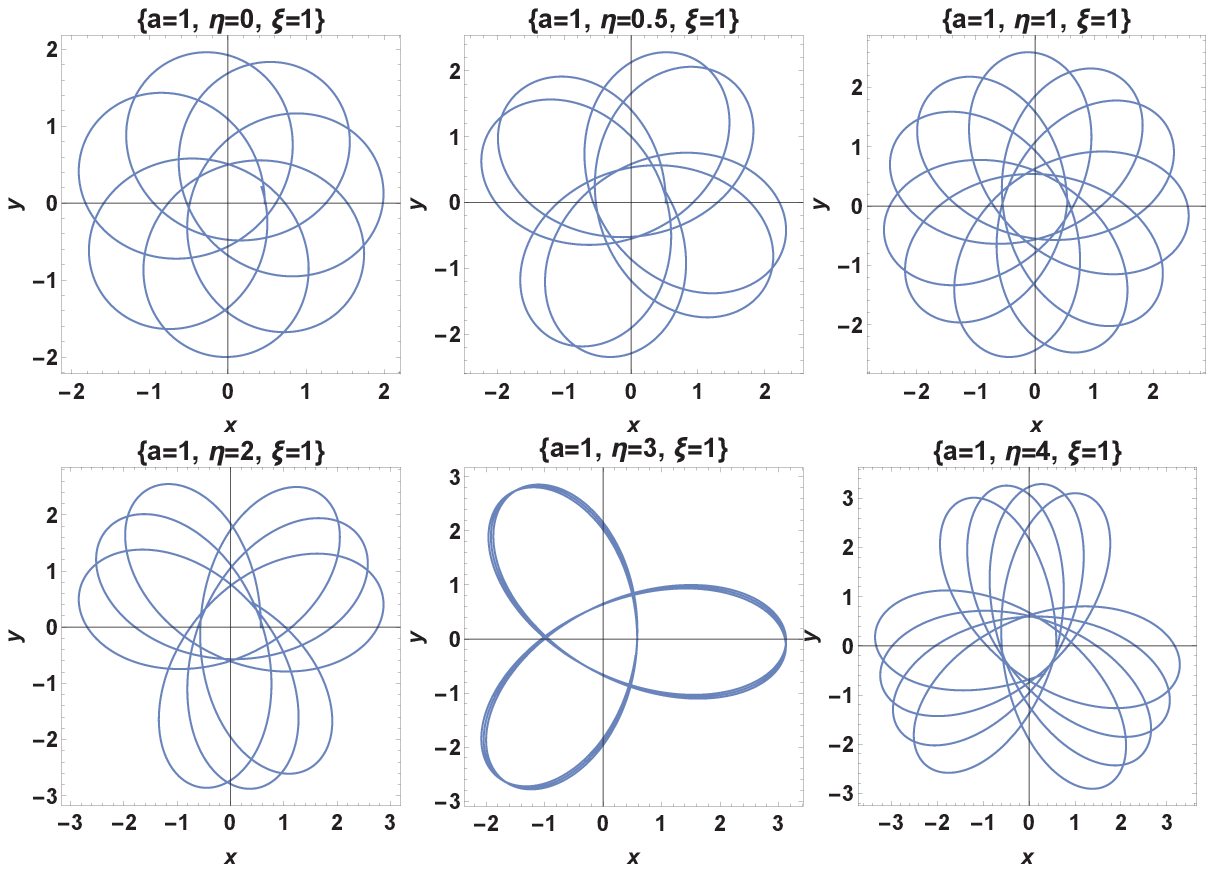}
\caption{The projective $x-y$ plane of the spherical photon orbits around the deformed Kerr black hole with the given initial angular momentum $\xi=1$ for a different deformation parameter $\eta$.  Here we set $M=1$. \label{fig5}}
\end{center}
\end{figure}
\begin{figure}[ht]
\begin{center}
\includegraphics[scale=0.8]{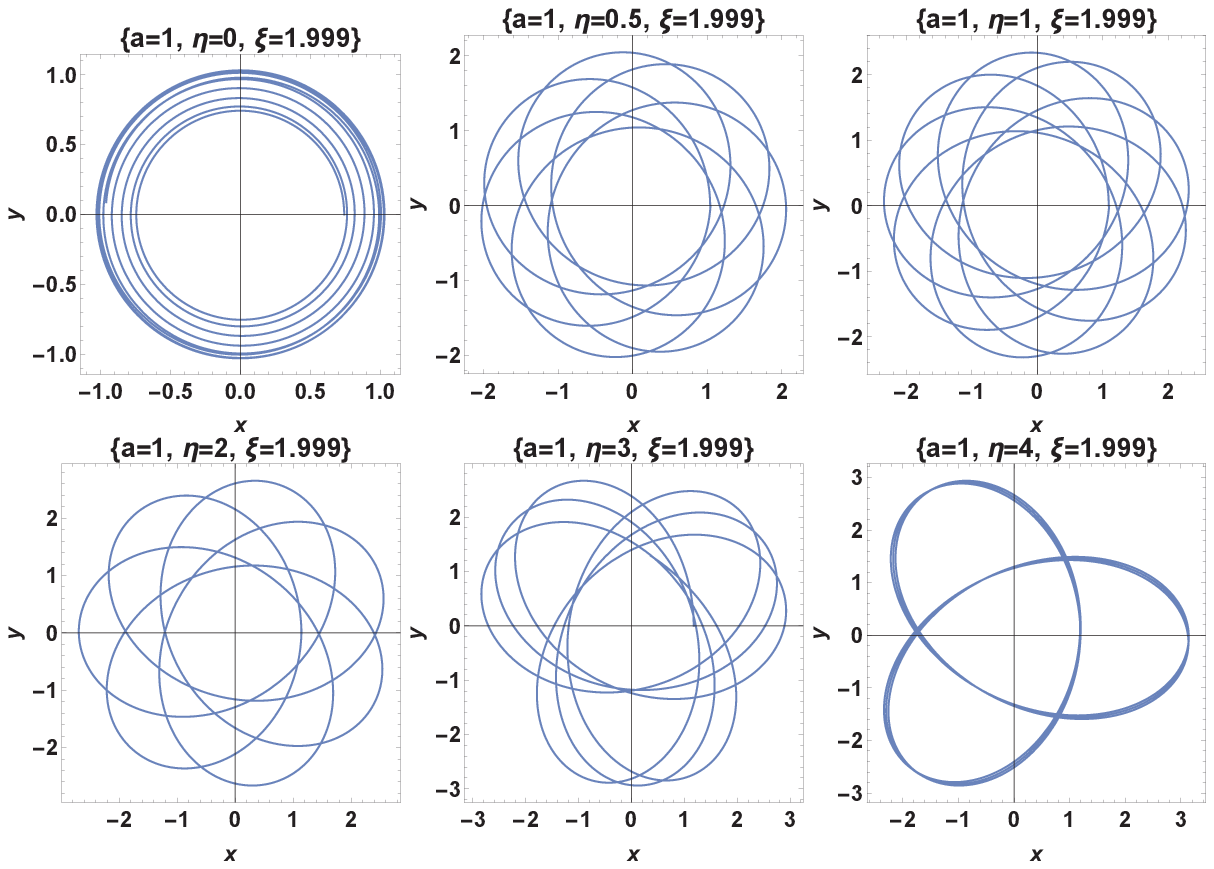}
\caption{The projective $x-y$ plane of the spherical photon orbits around the deformed Kerr black hole with the given initial angular momentum $\xi=1.999$ for a different deformation parameter $\eta$..  Here we set $M=1$. \label{fig6}}
\end{center}
\end{figure}

\section{Examples of Selected Spherical Photon Orbit Around the deformed Kerr black hole }
In this section, we shall present explicit examples of spherical photon orbits
around the deformed Kerr black hole. In Ref\cite{teo}, two prograde($\xi>0$ and $r_1<r<r_3$) and four retrograde($\xi<0$ and $r_3\leq~r<r_2$) spherical photon orbits around extreme Kerr black hole was plotted out, including two special case: a zero-angular momentum($r=r_3$) photon orbit and one orbit at initial radius at $r=r_{\dot{\varphi}}$ where $\dot{\varphi}=0$.
To compared with the result of the extreme Kerr black hole, we also plot the spherical photon orbits
around the deformed Kerr black hole($a=M$).
These orbits can only be obtained numerically, by integrating the first-order differential
of Hamiltonian formulation.

Firstly, we plot the same Teo\cite{teo}'s examples of the three-dimensional($x-y-z$) spherical photon orbit around the  extreme Kerr black hole
at the initial  angular momentum $\xi=0, -1, -2, -6, 1, 1.999$, respectively. Besides
the three-dimensional spherical photon orbit in cartesian coordinates, We also plot the projective plane of $x-y$ and $x-z$
, $\theta-\varphi$ of the spherical photon orbit around the extreme Kerr black hole to more vividly illustrate the properties of
periodic latitudinal oscillations. This six examples of the spherical photon orbit is shown in Figs. \ref{figure3} and \ref{figure4}. Notice that the periodic motion is mainly reflected on the $x-y$ projective plane.  From this pictures, we get the following interesting properties of the orbits :(1) The $x-y$ plane orbit of the zero-angular momentum($r=r_3=1+\sqrt{2}$ and $\xi=0$) looks like
 four-leaf colliding in the center.
(2) The $x-y$ plane orbit with initial value ($r=r_{\dot{\varphi}}=3$ and $\xi=-2$), where $\dot{\varphi}$ changes sign, is Periodic cuspy orbits. The photon is moving vertically whenever it is at the equator(see the correspondent $x-z$ plane). This behavior can
be understood from the Lense Thirring effect: the dragging of inertial frames is
strongest at the equator, and in this case, it precisely cancels out the retrograde
motion of the photon. Away from the equator, the dragging becomes weaker and
so the orbit regains its retrograde character\cite{teo}. (3) The $x-y$ plane orbit with the initial value ($r=1+\sqrt{3}$ and $\xi=-1$)  is Trochoidlike trajectory. This photon orbits do not have a fixed azimuthal direction.
(4) The $x-y$ plane orbit with the initial value ($r=1+2\sqrt{2}$ and $\xi=-6$) looks like a pancake around by sixteen stripes. (5) The $x-y$ plane orbit with the initial value ($r=2$ and $\xi=1$) looks like precession of circles.  (6) The plane orbit with the initial value $r=1.0316$ and $\xi=1.999$ is one latitudinal oscillation and its  $x-y$ plane orbit  is spiral circle and a
helical pattern. (7) The value of $\theta$ is periodic oscillated with the change of $\varphi$ for all of this spherical photon orbits.

To compared the spherical photon orbit around the deformed Kerr black hole with the case of the extreme Kerr black hole\cite{teo}, we
also plot six example of the projective $x-y$ plane of the spherical photon orbits around the deformed Kerr black hole  in Figs. \ref{fig1},\ref{fig2},\ref{fig3},\ref{fig4},\ref{fig5} and \ref{fig6}.
Especially,  when the deformation parameter take the value $\eta=0.5,1,2,3,4$,  the shape and periodicity
of latitudinal oscillating orbit is great different from the case of Kerr black hole and exhibit qualitatively different behavior for the given initial angular momentum $\xi=0, \xi_{\dot{\varphi}},-1, -6,1, 1.999$, respectively.  The illustrations of these examples
manifest that spherical photon orbits around the deformed Kerr black hole have a variety of interesting behavior that are absent in circular
orbits. These examples, in principle, may provide a possibility to test how astronomical black holes with the deformation parameter $\eta$ deviate from the Kerr black hole.

\section{summary}

 In this paper, we
have studied the spherical photon orbits around the deformed Kerr black hole. The change in the azimuth $\Delta\varphi$  and the angle of dragging of nodes per revolution $\Delta\Omega $ of a latitudinal oscillated orbit is calculated analytically. Besides
the three-dimensional spherical photon orbit in Cartesian coordinates, we plot the projective plane of the three-dimensional orbit
and the $\theta-\varphi$ plane of the spherical photon orbit around the extreme Kerr black hole to more vividly illustrate the properties of
periodic latitudinal oscillations.  This spherical photon orbits shown in Figs. \ref{figure3} and \ref{figure4} are latitudinal oscillations and periodicity. Finally, we also illustrate different behavior with six representative examples of such orbits around deformed Kerr black hole to compared with the case of Kerr black hole in Figs. \ref{fig1}, \ref{fig2}, \ref{fig3}, \ref{fig4}, \ref{fig5} and \ref{fig6}, including two special case: a zero-angular momentum($r=r_3$) photon orbit and one orbit at initial radius at $r=r_{\dot{\varphi}}$ where $\dot{\varphi}=0$. As we have seen,
these orbits exhibit a variety of interesting behavior that are absent in circular
orbits.

\section{\bf Acknowledgments}

Changqing's work was supported by the National Natural Science
Foundation of China under Grant Nos.11447168. Chikun's work was
supported by the National Natural Science Foundation of China under
Grant Nos11247013; Hunan Provincial Natural Science Foundation of
China under Grant Nos. 12JJ4007 and 2015JJ2085. I would like to thank Carlos A. R. Herdeiro and Pedro V. P. Cunha
for reading the manuscript and for their useful comments.

\end{document}